\documentclass[pdflatex,twocolumn,a4paper,showpacs,10pt]{revtex4-1}
\usepackage[utf8]{inputenc}
\usepackage[english,spanish]{babel} 
\usepackage[T1]{fontenc}
\usepackage{amsfonts}
\usepackage{amsmath}
\usepackage{amssymb}
\usepackage{graphicx}
\usepackage{dcolumn}
\usepackage{latexsym}
\usepackage{hyperref}
\usepackage{bm}
\usepackage{color}

\begin{document}

\title[PTP-Au-NP]{Microscopyc description of the Photothermal increase of temperature for metallic nanoparticles excited with short-laser pulses}

\author{M. Rodríguez-Matus}

\address{ Posgrado Ing. Elect., Universidad Aut\'onoma de M\'exico, Circuito Exterior Universitario, C.P. 04510 Coyoac\'an. Mexico City, Mexico}
\affiliation{ Instituto de Ciencias Aplicadas y Tecnología, Universidad Nacional Autónoma de México, Circuito Exterior S/N, Ciudad Universitaria, 04510, Mexico City}

\author{C. Garc\'ia-Segundo} 
\address{ Instituto de Ciencias Aplicadas y Tecnología, Universidad Nacional Autónoma de México, Circuito Exterior S/N, Ciudad Universitaria, 04510, Mexico City}

\author{Jean-Patrick Connerade}
\address{ QOLS Group, Physics Department. Imperial College London: SW7 2BW, UK}

\date{ \today}

\begin{abstract}
The pulsed photothermal phenomenon due to optically absorbed energy, result from non-radiative decay mechanisms, which in nature, these imply the temporal change in the local free energy and thus a temporary change in the local temperature. At the nanoscale, this is a prediction broadly described in terms of macroscopic variables. Here we introduce a formalism based on the Jarzynski's physical statistics description, for interpreting the equilibrium free energy difference between two configurations of a metallic nanoparticle. In this way, within a finite-time span, we describe the temporal increase of the local free energy and thus of the local temperature arising from temporarily bringing the sample far from thermal equilibrium. The result is an expression for which one can get the photothermally induced change of temperature for a metallic nanoparticle. For practical purposes, we limited the study to Au nanoparticles. The study is made as function of the particle size and the optical properties for wavelengths spanning the optical range. The assessment indicates that, for nanoparticles with radii shorter than 40 nm, the temperature change is strongly dependent on particle size and on the illumination wavelength. While, for near 40 nm particle radii, the current description and the known formalism, based on macroscopic variables, predict the very same temperature change. At the closing we discuss additional possible thermodynamic consequences associated to the scale considerations.
\end{abstract}


\vspace{2pc}


\maketitle

\section{Introduction}
Equilibrium situations, which are fundamental in thermodynamics as they allow basic quantities such as temperature, entropy, etc. to be defined. However, they generally correspond to idealised situations. For example, even systems as stable as the atmospheres of the sun and stars cannot be described using the full apparatus of equilibrium thermodynamics. They require the introduction of local thermodynamic equilibrium (LTE) which allows energy to escape from the medium by radiation. As a consequence of this relaxation of full equilibrium conditions, different temperatures exist for ions and for electrons, under the assumption of two separate Maxwellian velocity distributions. Similarly, in studying molecular spectra it is a familiar and useful step to introduce different temperatures coexisting within the system for different kinds of transition (electronic, vibrational, rotational, etc.) Nonetheless, the overall architecture of equilibrium  thermodynamics serves as the conceptual guide, the key aspect being to identify changes needed in the theory to account for departures from full equilibrium.

The advent of pulsed lasers and the interest in the interactions they produce in dense media emphasize the need to extend the study of departures from equilibrium. Here again, a fruitful first approach is to start from the equilibrium equations and explore modifications which can help represent 'real' situations with the help of comparisons with experimental data. The aim is to extract information concerning temperature changes and, possibly, other thermodynamic variables by identifying which equilibrium conditions can be relaxed in order to describe, for example, the interaction between a non-destructive laser pulse and a thermoelastic solid. Assumptions must be made concerning the preservation of the nature of the target (for example : no phase change is induced by the laser pulse).

Within this context is that we discuss the photothermal (PT) effect, which in turn correspond to the local change of temperature that is induced by local absorption of electromagnetic radiation, within a nanoscale solid sample. When the related solid is a metal at the scale of nanoparticles,n the PT effect occurs with rather high efficiency. Currently, reports indicate that such efficiency is function of the particle geometry, absorption cross section, volume, surface area and density. These conditions combine the strong confinement of Mie resonant and non-resonant absorbed electromagnetic field, which then gradually decays through ohmic losses, the type of collisional processes fueling non-radiative decays; i.e. Joule heating \cite{BaffouPRB2010, JPC2017, ElSayed, HillenbrandNature, JiangJPCC2013}. 

On these scales, it is considered that the PT effect in the nanoparticle produce an temperature increase ($\Delta T$) that depends upon the absorbed optical power density ($W_{\alpha}$) and on the particle's thermal conductivity ($\kappa$); hence, it is controlled by the laser intensity and the pulse-width, $\tau$, respectively. As result, the temperature rise is expressed as $\Delta T = W_{\alpha}/[4 \pi \kappa r_o]$. With the help of this expression several authors provide a close quantification of the PT related change of temperature \cite{Richardson, JiangJPCC2013, Selmke_ThermalLens}, and thus to better understand its importance and physical implications \cite{AngerPRL2006, HotBrownianMotion, BraunCichios2014, Desiatov2014, HillenbrandNature}. 

However, when the source of illumination are short laser pulses  ($ \sim ns$), interacting with metallic nanoparticles in the Mie regime ($r_0 \leq 50 nm$), the macroscopic temperature description is at the borderline of the thermodynamic limit \cite{HartmannPRL2004, JPC2017, CGS2010}. Therefore since $\Delta T$, is described in terms of macroscopic parameters, thus it is suggestive that it requires to be re-casted to obtain an expression in which the micro- or rather nanoscale, can be better described. The task we cover here, is to introduce a formalism to get an expression for the temperature change out of the PT phenomena in a metallic nanoparticle ($r_0 \leq 50 nm$), illuminated by $ \sim 7~ ns$ laser pulses. From such result, as we describe, several other consequences arise. The clear interpretation of this parameter is a current subject of intense research; specially after the fast-growing fields of so-called thermoplasmonics, and plasmonic photothermal therapy (PPTT)applications \cite{BaffouPRB2010, ElSayed, ElSayed2011}, among many other areas of interest. 

As starting point, we recall that, like in macro-scale, at nanoscale, the thermal equilibrium condition posses a central role in the interaction of radiation with matter. In this regard, its consideration addresses the longstanding question: for how small a system is it reasonable to introduce the concept of temperature in the classical sense? In other words, how can the classical temperature be properly defined for a system verging on the microscopic? In part these questions are broadly reviewed in \cite{JPC2017}, while they were the subject of discussion in \cite{HartmannPRL2004} and \cite{LebowitzPRL1969}. In \cite{HartmannPRL2004}, the authors managed to describe how a system obeying statistical mechanics leads to proper macroscopic thermodynamics (scale-up). In the later, the authors use the framework of statistical mechanics to introduce a description of temperature on the nanoscale, including a criterion to identify its range of validity.

 In the photothermal phenomenon (PTP) in nanoparticles (NP's), induced by short laser pulses ( $10 ~ns$ time scale), it is implicit that, during the interaction process, locally the sample is driven away from equilibrium for a finite-time span. As described in \cite{JarzynskiPRE1997, JarzynskiJSP, JarzynskiPRL1997}, this can be interpreted as an induced free-energy difference between the initial and the final equilibrium. However, this time limited to the PTP, where the free-energy change is due to the portion of pulsed optical energy that is absorbed within the sample's volume and then converted into heat via non-radiative processes, limited to the interaction volume and without prompting chemical reactions. Because the pulsed excitation, the change in the local free energy is temporary in nature \cite{FrohlichBook}; which in itself represent the temporary increase of the local internal energy density  ($\rho_e(t)$). Let this difference of free energy be represented in terms of the time dependent partition function $Z(t)$, such that its temporal change shall equate with the temporal change of the energy density \cite{JarzynskiPRE1997, JarzynskiJSP, JarzynskiPRL1997}. This interplay was discussed in \cite{CGS2009, CGS2010}, where the relationship between the optical thickness and the thermal thickness, was explicitly displayed and justified. In there, it was made apparent that the coupling of optical energy with the thermal energy depends from two physical properties alone: the optical absorption and the thermal conductivity (the transport process). However the description, placed for macro-scale, made use of a time independent partition function, and served to proof how to obtain the optical properties of thin films out on non-radiative measurements. Since then, it was suggested that as the sample's scale shortens, the time dependent case would apply, and explicitly it was identified that a $Z(t)$ type function would be required. Related to the PTP description, it also, in \cite{CGS2009, CGS2010} made apparent how is that the time evolution of $\rho_e(t)$ shall equate with the product between the thermal conductivity($\kappa$), the optical absorption($\alpha$) and the temperature change ($T_l = T_0 + \delta T$), all rated by the depth of the interaction, $z_1$. Therefore, departing from the description in \cite{GS2010}, this time using the time dependent partition function $Z(t)$, we consider to analyze the interaction of a metallic particle with radius at nanometer scale, say $r_0 \leq 50 ~nm$ and \textit{7 ns} laser pulses. Thus the known analytical description must be modified to be expressed as 

\begin{equation}\label{TG_1}
\frac{d\rho_e}{dt} =\frac{d}{dt} \left[-k_B T \log(Z(t))\right]=\frac{\kappa\alpha T_l}{z_1} ~.
\end{equation}

 Let us consider an equilibrium initial state at temperature $T_0$, with $F_0$ the sample's free-energy in absence of the external field, and thus $Z_0$ as the initial state. Eq. (\ref{TG_1}) implies that each pulse of external field shall perform an amount of work equal to the change of free-energy $\Delta F$, bringing the sample to a temperature $T_l$, implying the change of temperature $\delta T = T_l - T_0$. If we consider the dimensions of the nanoparticle, then it is required to scale the problem so as to identify the limit within which the existence of a local temperature is being assumed. Specially on the view that as the size of the nanoparticle decreases, the thermal conductivity and the partition function cannot be calculated as in a continuous medium; see \cite{GemmerBook2009} for the related details. From the scope of the finite time of interaction, one can infer that the change exhibited by the partition function is limited by the time span of interaction of the radiation with the nanoparticle. As the volume and the number of constituents in the nanoparticles remain constant with its energy as a random variable, then it is appropriate  for this system to be described by a time dependent canonical partition function. It turns out that the problem can be treated by means of the so-called Jarzinski equation, which provides a way to relate energy distribution between the work and heat produced within a time interval, to the change in free energy, $\Delta F$. Its analytic content,

\begin{equation}\label{Jarzynski1}
\overline{e^{-\beta W}}=e^{-\beta\Delta F}=\frac{Z(t)}{Z(0)},
\end{equation}

describes the time-dependent states distribution, with respect to an initial $Z(0)$, in terms of an ensemble average independent of: (a) the trajectories for the increase of free energy and (b) time. Thus, it is enough to know two equilibrium states (initial and final), to obtain the non-equilibrium value at any time. However quasi-equilibrium conditions and sufficiently weak coupling between the system and the reservoir conditions must apply. 

 According to the Wiedemann-Franz law \cite{PeierlsBook2001}, the thermal conductivity of a metal is proportional to the temperature and to its electrical conductivity. However as a redefinition of the partition function is required when the nanoparticle size is reduced, the thermal conductivity needs to be recast in terms of the internal energy and the number of the confined atoms, determined statistically by a Boltzmann distribution. Thus, the thermal conduction implies diffusivity, due to occur within the time span $\tau_{ext} < t < \tau_\delta $; here $\tau_{ext} = 1/ \delta \epsilon$ is the extinction time and $\tau_\delta = 2 \pi n / \delta \epsilon$, with $\delta \epsilon$ the energy bandwidth within which the nanoparticle excitation occurs \cite{GemmerBook2009} and $n$ the number of atoms forming the nanoparticle. For the sample constituents, onside of the time constraints and for a system close to equilibrium, we can assume that the probability distribution of the components is approximately canonical, and that the energy diffusivity can be linearly approached by Fourier's law. This broad scenario has been discussed in \cite{GemmerBook2009}, resulting in an appropriate expression for the thermal conductivity,

\begin{equation}\label{heatconductivity}
\kappa ~ = ~\frac{2\pi \lambda ^2 n}{\delta \epsilon} \left( \frac{\Delta E}{T}\right)^2\frac{ne^{-\frac{\Delta E}{T}}}{\left( 1+ ne^{\frac{\Delta E}{T}}\right)^2} ~~;
\end{equation}  

then, it turns out that $\kappa$ is function of the particle size through the value of $n$. 

In order to describe the change in local energy distribution as function of the optical absorption of the nanoparticle, it is necessary to implement Mie's theory to obtain the absorption cross section, that calculates the total amount of radiation absorbed by the nanoparticle. Also, as the energy distribution is proportional to time and irradiance, we assume that the excitation (the increase of temperature) occurs only while the radiation pulse is interacting with the nanoparticle. 

These considerations are already contained in equation (\ref{TG_1}). Thus, from time integration within the pulse time-width ($t_0$), and substituting (2) and (3) into (1) after defining

\begin{equation}
\gamma= \Delta F \kappa\alpha/k_Bz_1\beta,  
\end{equation}

we get a statistically based expression for the temperature increase of a given nanoparticle type,

\begin{equation}\label{final}
\delta T=T_0 ~~\frac{\log(Z_0)}{\log(Z_0)- \Delta F \beta t}  ~\left\lbrace \left[\log(Z_0-\Delta F \beta t_0)\right]^{\gamma}-1\right\rbrace~~.\\
\end{equation}

This equation is simplified by Taylor expansion, to get

\begin{equation} \label{T}
\approx T_0\left\lbrace \frac{\log(Z_0)}{\log(Z_0)- \Delta F \beta t}\left[1-\frac{\kappa\alpha}{k_Bz_1}\frac{t_0}{\log(Z_0)}\right]-1\right\rbrace. 
\end{equation}

This solution relates the change of temperature experienced by the nanoparticle to: its capacity to absorb optical energy, the way the heat is conduced through the medium and the way the energy is distributed within the system (the partition function). It also depends on the time of interaction between the external field and the nanoparticle and varies inversely to the size of the spherical particle. In other words, from the statistical estimates, it is apparent that the temperature increase is a function of particle size.


In order to exhibit the temperature change that a nanoparticle experiences while illuminated by a laser, we studied equation \ref{final} varying the main parameters: the wavelength of the incident field and the radius of a gold nanoparticle. Here we report the analysis for nanoparticles with radius between 2 and 90 \textit{nm}, illuminated with wavelengths between 300 and 1200 \textit{nm}. The maximum intensity was obtained using $I_0=2P_h/\pi\omega_{0,h}^2$ \cite{SelmkeJOSAA2012} where $P_h=22500J$ is the laser power and $\omega_{0,h}^2$=0.330$\mu$m the beam waist. The initial temperature is assumed to be 20$^\circ$C. The absorption cross section was calculated via Mie's theory, $W_{abs}=\sigma_{abs}I_0$, and the refractive index for the surrounding medium was taken from the literature  and the gold refractive index is calculated by applying the equation $Q_{size}=Q_{bulk}\left[1-\exp\left(R/R_0\right)\right]$ which introduces a correction due to the size of the system \cite{Prashant, Bohren, ScaffardiNanoIoP2006}. The number of particles that compose the nanoparticle is obtained by $n=30.89602D^3$, with $n$ the number of atoms and $D$ the particle diameter \cite{Noatomos}. 

As proof of consistency of the analytic results, we produce numerical calculation sto compare the temperature change $\delta T$ calculated using (\ref{T}), with that $\Delta T $ calculated as reported in the literature \cite{BaffouPRB2010, JiangJPCC2013, Selmke_ThermalLens}. The optical absorption was fitted by numerical calculations from \cite{ScaffardiNanoIoP2006}, and compared with the experimental curves in \cite{Prashant}. Notice that, in either case, the temperature difference exhibited in metallic nanoparticles is proportional to the amount of absorbed energy (or rather the absorption cross section $\sigma_{abs}$), and varies inversely as the size of the nanoparticle. In  (\ref{final}), the particle size is implicit from the statistical distribution, by means of (\ref{TG_1}) and (\ref{heatconductivity}).

\begin{figure}[h]
\begin{center}
	\includegraphics[width=0.45\textwidth]{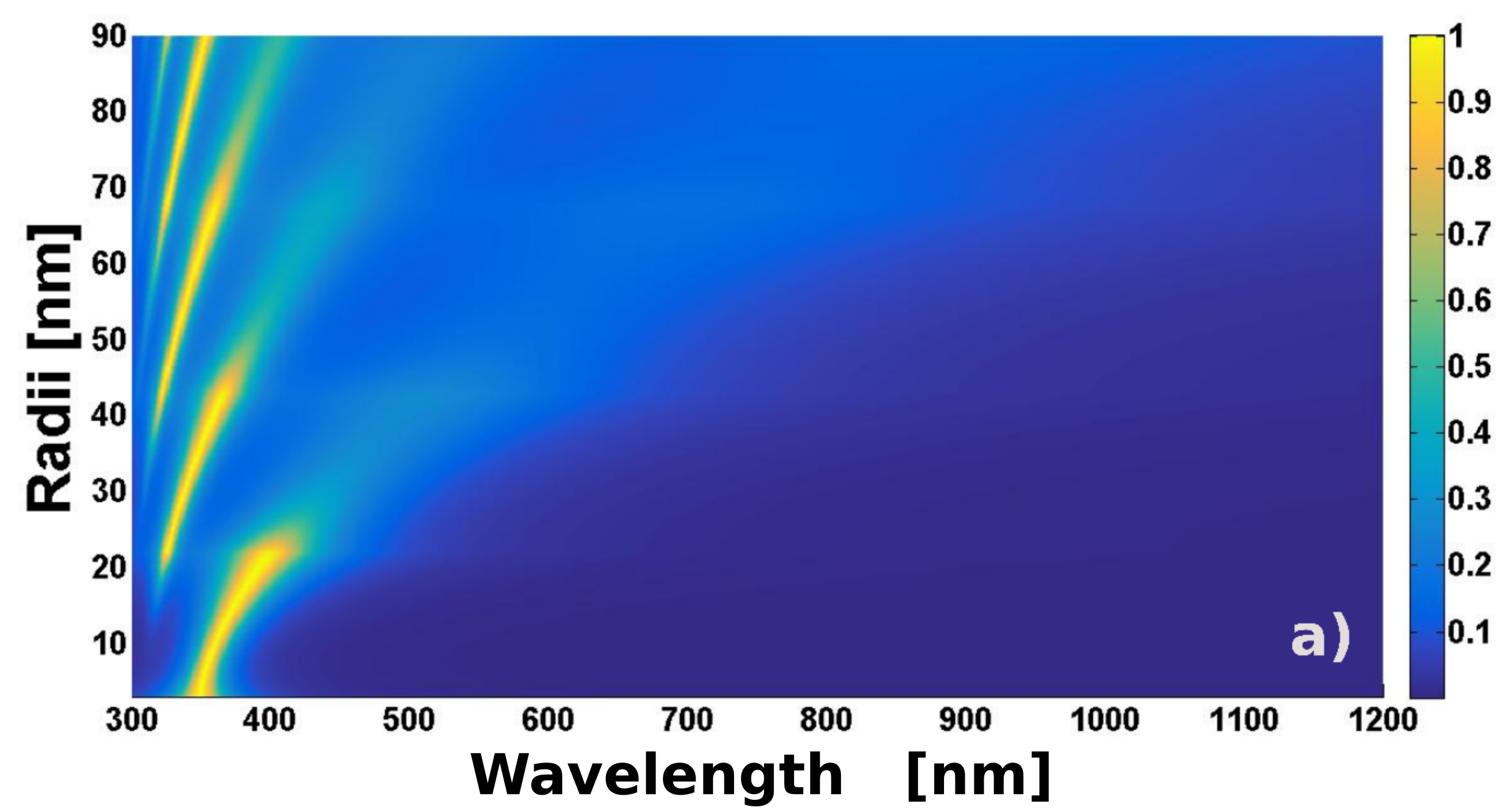}
\includegraphics[width=0.45\textwidth]{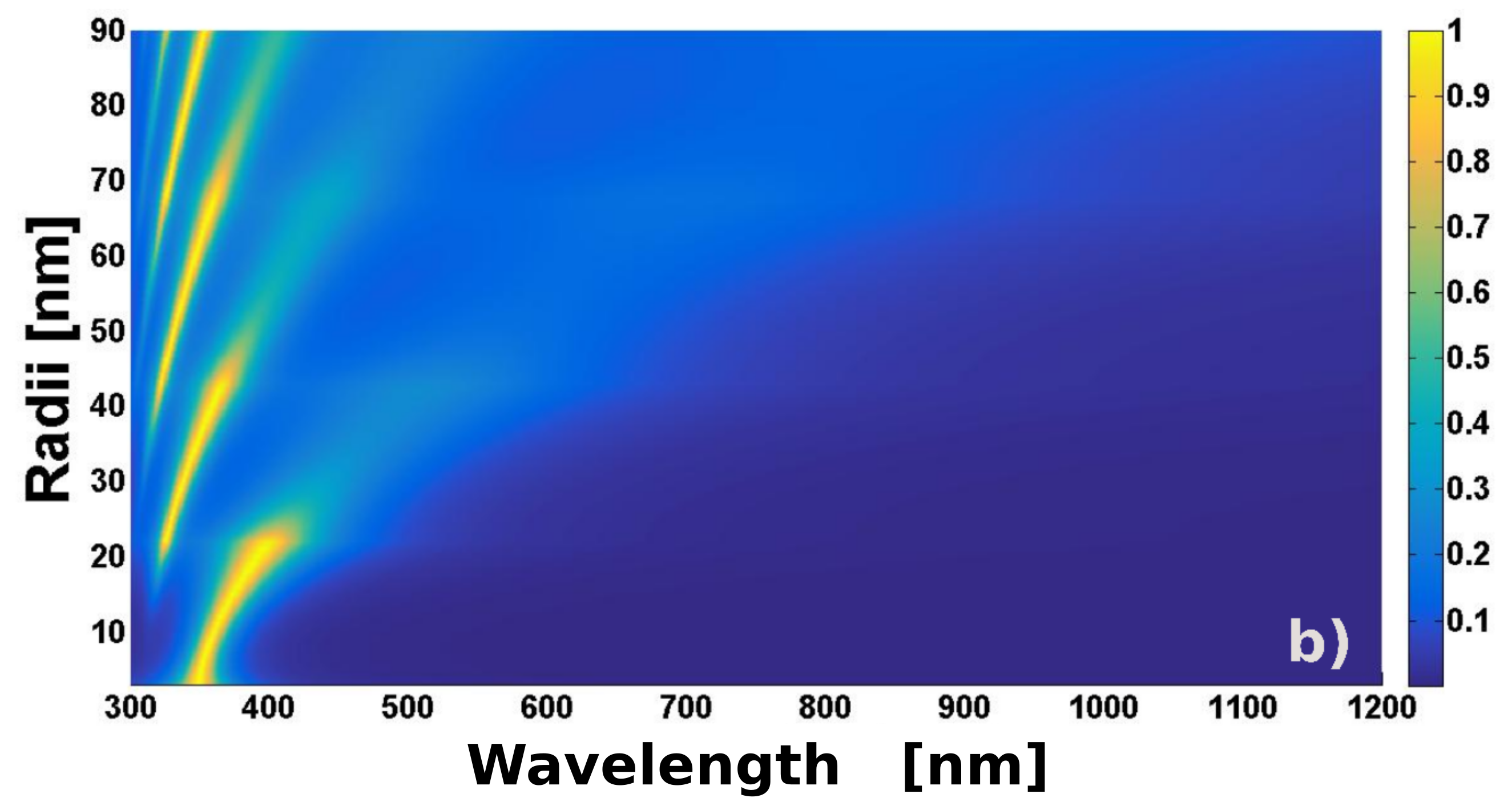}
	\caption{Normalized temperature distribution map, as function of: the particle radii ranging from 2 to 90 \textit{nm}, and for illumination wavelength from 300 to 1200 \textit{nm}. (a) is $\delta T$ estimated from \ref{T}); (b) is for $\Delta T $, as in \cite{BaffouPRB2010, JiangJPCC2013, Selmke_ThermalLens}} \label{Fig1}
\end{center}
\end{figure}

The comparison is displayed in Fig. (\ref{Fig1}), as a color map of the normalized temperature distribution for gold as function of the nanoparticle radii, and wavelengths from 300 to 1200 \textit{nm}. Phenomenologically the mismatch is apparent between the current calculation using \ref{T}, and that using $\Delta T$ as in \cite{BaffouPRB2010, JiangJPCC2013, Selmke_ThermalLens}. Noticeably, and regardless of the nanoparticle radii, the PT effect with larger temperature changes,'hot modes', occur for excitation wavelengths shorter than 500 nm. The normalized temperature scale is introduced to better compare the thermal distribution resulting from the numerical calculations out of the statistical and macroscopic variable theoretical descriptions, respectively.

\begin{figure}
\begin{center}
	\includegraphics[width=0.4\textwidth]{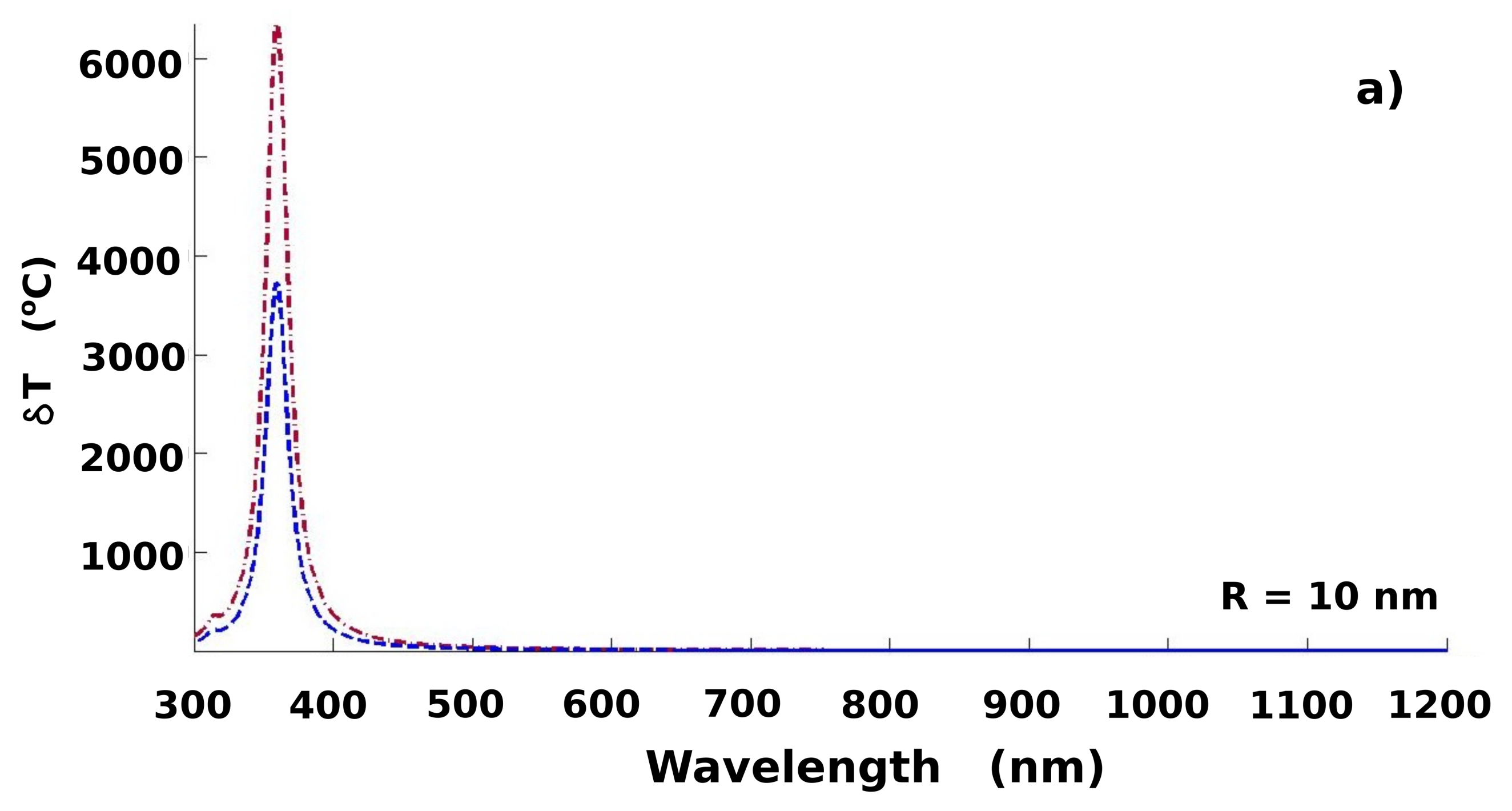}
	\includegraphics[width=0.4\textwidth]{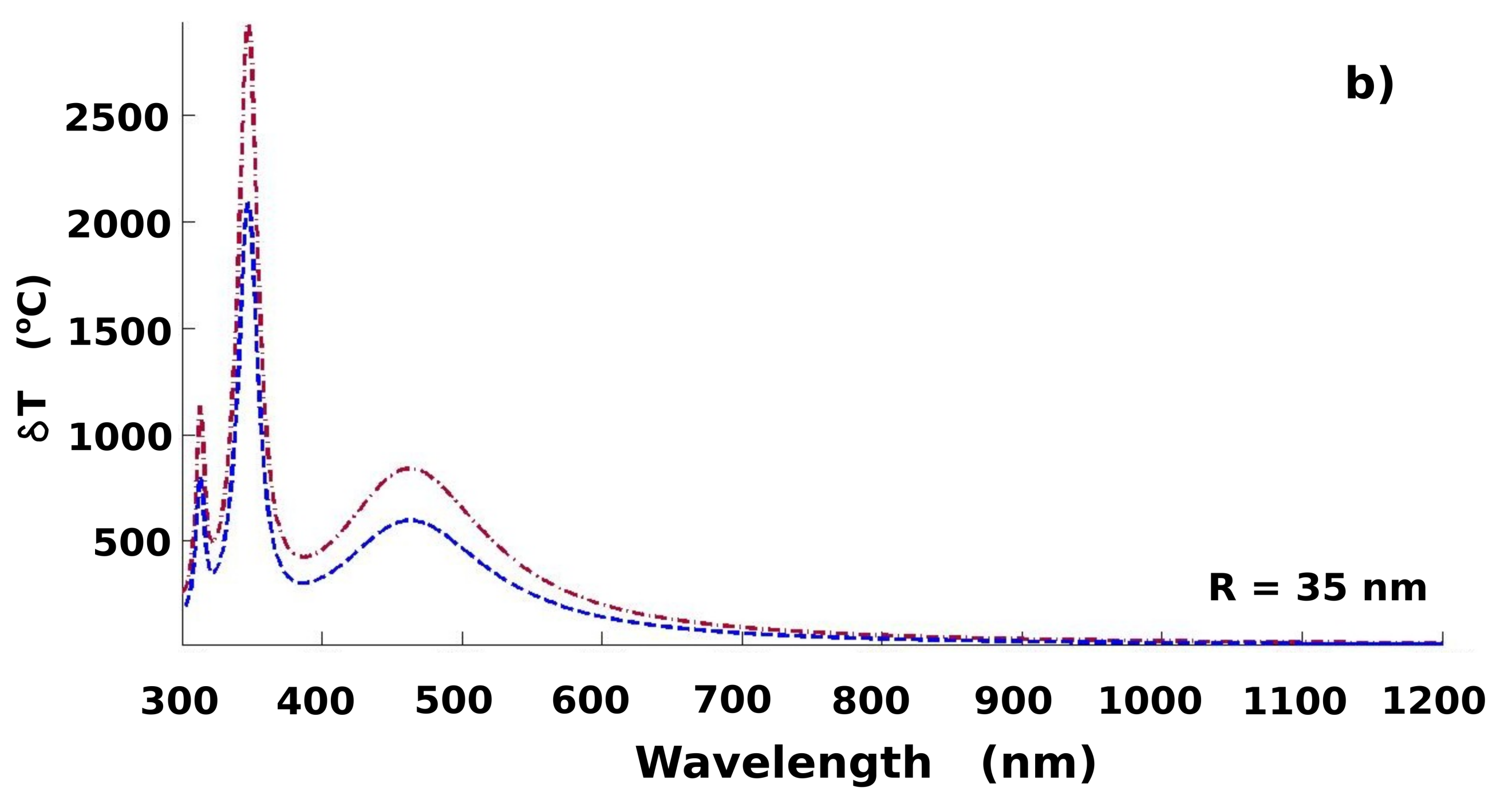}
	\includegraphics[width=0.4\textwidth]{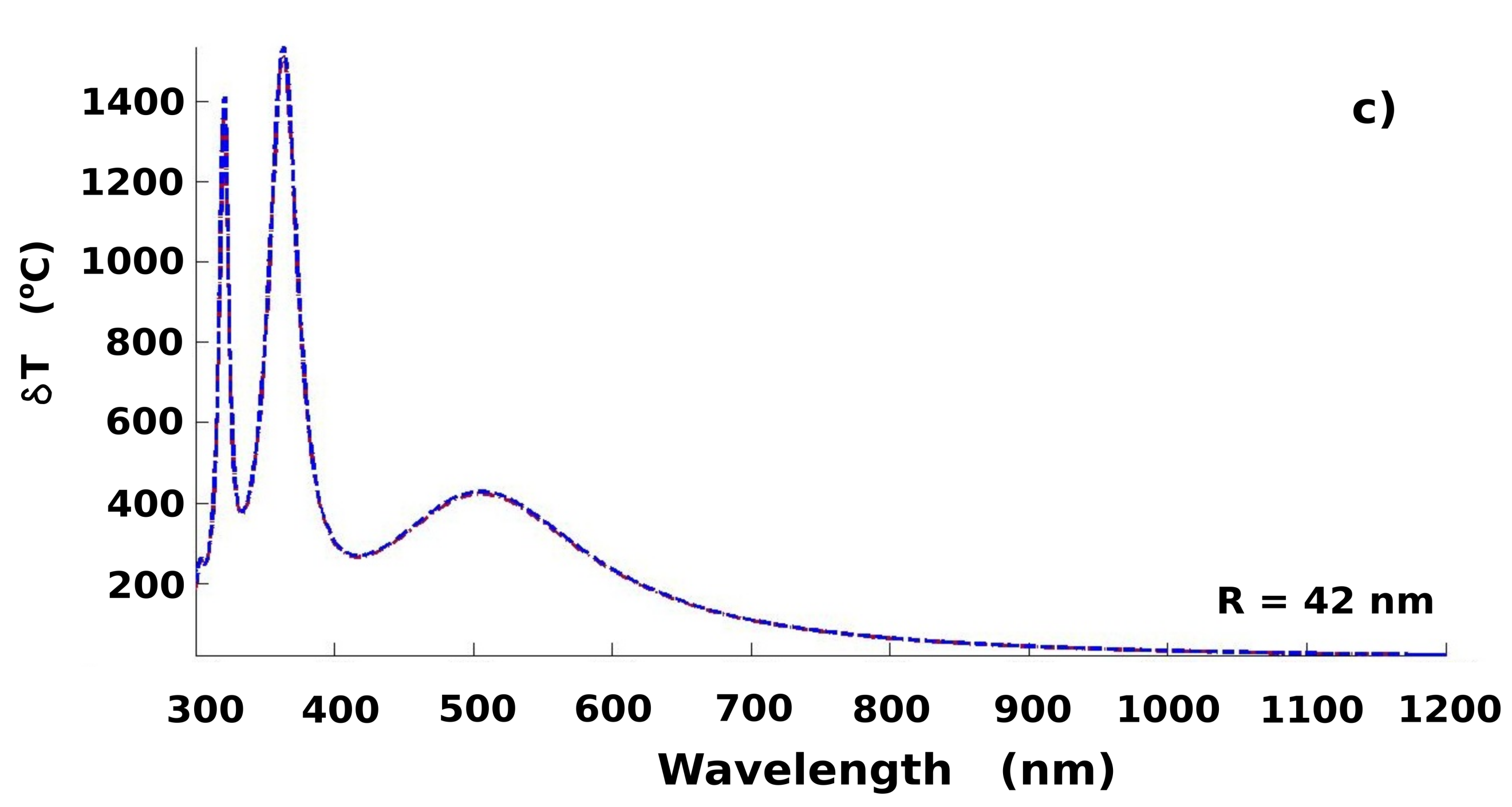}
	\includegraphics[width=0.4\textwidth]{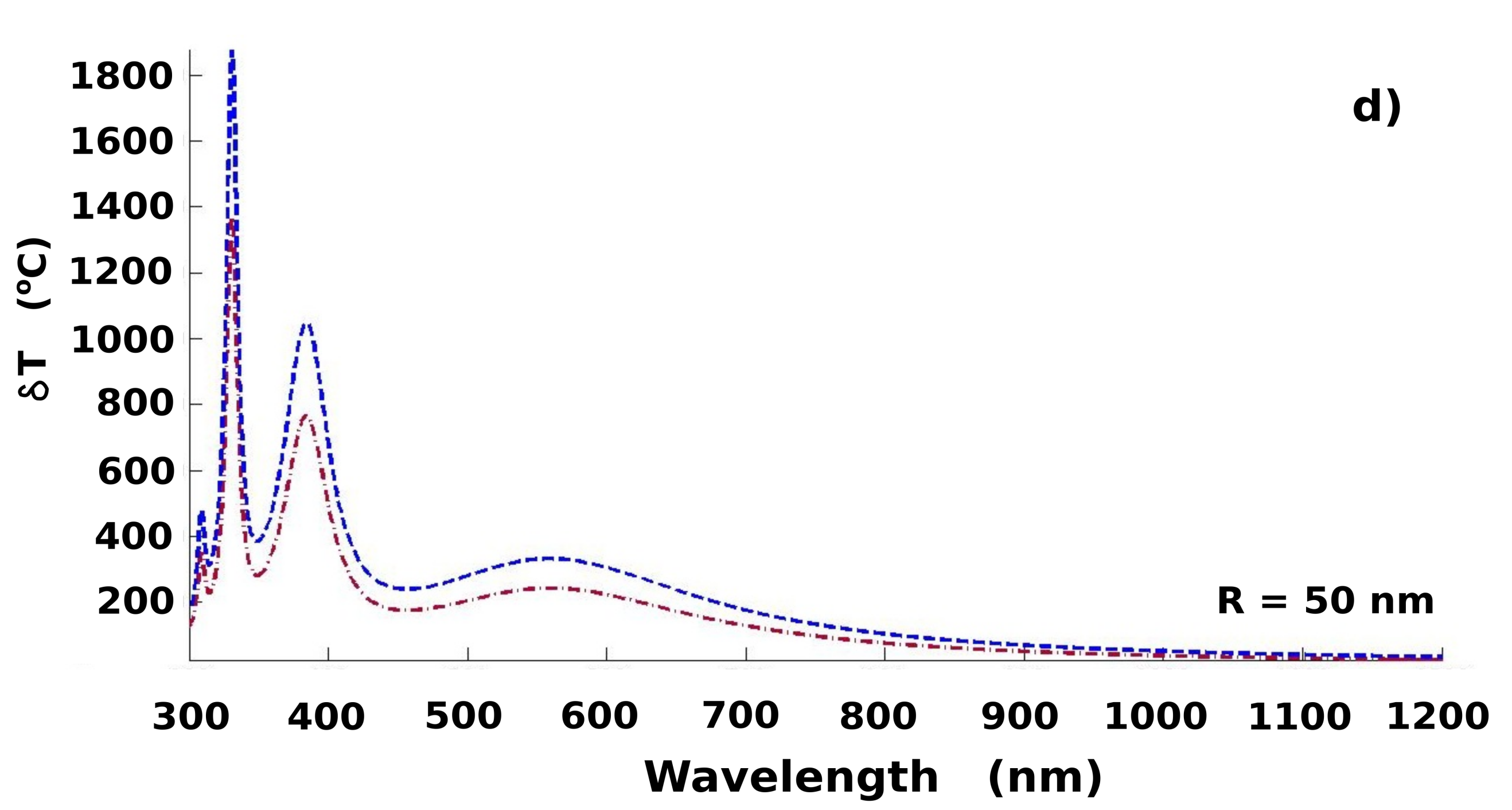}
	\caption{Temperature rise as function of the wavelength for the nanoparticle radius:(a) 10 nm, (b) 35 nm, (c) 42 nm and (d) 50 nm. The red line describes the current model estimate, and the blue line is for the estimate using $\Delta T $, as in \cite{SelmkeJOSAA2012, JiangJPCC2013, Selmke_ThermalLens}.} \label{Fig2}
\end{center}
\end{figure}

Next we ploted the temperature distribution, calculated from the current statistical analysis and $\Delta T$ as in the literature, as function of the illumination wavelength between 300 and 1200 nm,for nanoparticles of radii 10, 35, 42 and 50 nm. The result is displayed in the Fig. (\ref{Fig2}), showing curves whose the trajectories exhibit phenomenological closeness. In particular, for nanoparticles of 42 nm radius, both descriptions predict the very same response. However, for particles smalle that these 42 nm, and at wavelengths bellow the 500 nm, both descriptions display 'hot' modes with peaks at the same wavelengths; nonetheless, the current statistical analysis predict higher temperature change than that using macroscopic variables. Conversely, for NP's with radii larger than 42 nm, the statistical analysis predict lower values than these using macroscopic variables. Noticeably, the models disagreement, in terms of particle size and induced PT magnitude, seems to follow a match with the Mie limit, along with a strong dependency with the frequency bath in which the system is placed \cite{JPC2017, ElSayed}.

Specifically, from Fig. \ref{Fig2}a, it is shown that for a nanoparticle of 10 nm radius the temperature distribution calculated in statistical fashion, predict a temperature change near twice the one estimated by the literature description. However, when particle diameter is increased, the predicted changes of temperature converge until the radius equals 42 nm (see Fig. \ref{Fig2}c); at that stage the increase in temperature predicted by both models is the same. Besides, while the size is increased (above 42 nm) our model shows a smaller change in temperature in comparison with the existing model as shown in Fig. \ref{Fig2}d. We consider that, given the characteristics we considered, our model is not adequate to predict temperature where the system reaches the limit of 42 nm, where the discretization of energy levels no longer affects the photothermal effect, and as noted before, it has a close match with the Mie limit.

We also analysed the temperature distribution as function of the radius of the nanoparticle, when illuminated at the fundamental, second and third harmonic wavelengths of a Nd:YAG laser (1064, 532 and 355 nm, respectively). The results, displayed in Fig. (3), show us that for both models the increment in temperature do not exhibit close trajectories as displayed in Fig \ref{Fig2}. Neither the maximum and minimum values are located at the same wavelengths, as displayed en terms of wavelength for given particle size. The curves unveil that temperature coincidences occur around 42 nm, as noted before. At this value the plots cross over, predicting a higher temperature for smaller nanoparticles when the statistical analysis is applied and, a lower one if the nanoparticles are larger. The shift in the maximum of temperature, predicted by both models, can be related to the fact that the statistical model consider an extra feature that depends directly on the dimensions of the nanoparticle; i.e. the partition function. Of course, such consideration is out of scope in the model based on macroscopic properties. Anyway, it is significant that for larger wavelengths both approaches show consistent tendency for temperature increase with the particle size, whilst for short wavelengths the current model predict higher magnitude that the macroscopic variable estimate.

\begin{figure}[h]
\begin{center}
	\includegraphics[width=0.45\textwidth]{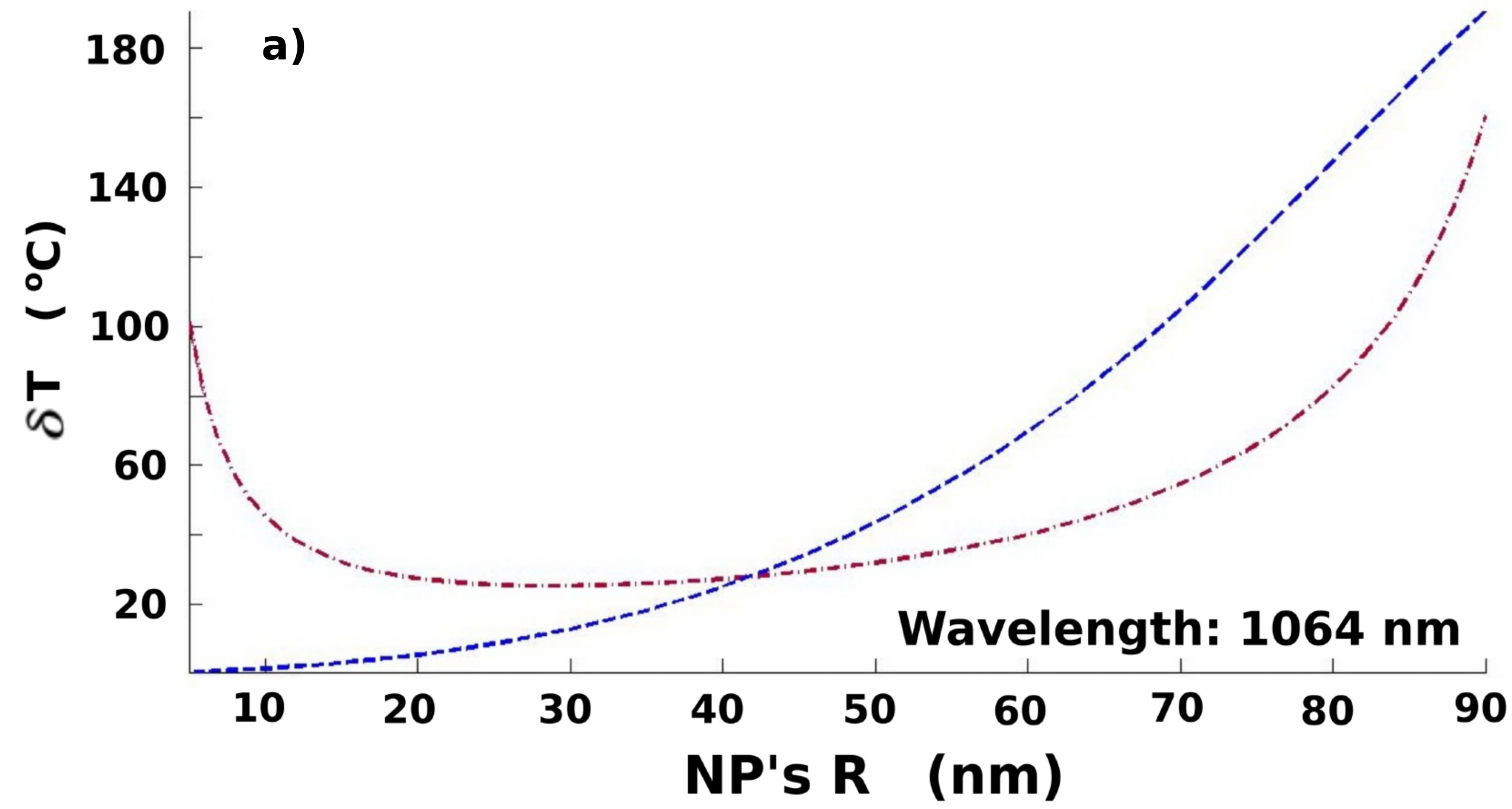}
\includegraphics[width=0.45\textwidth]{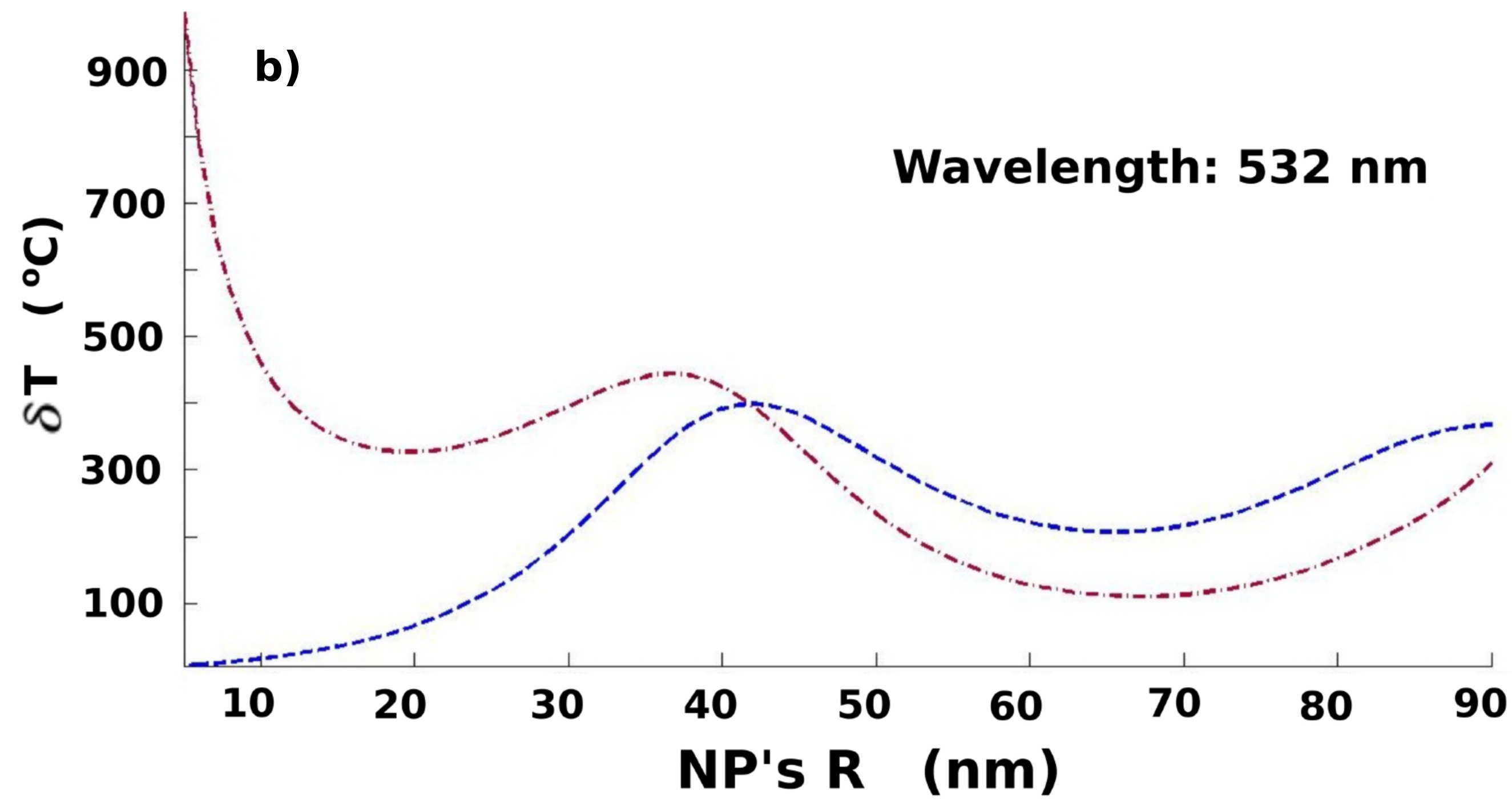}
\includegraphics[width=0.45\textwidth]{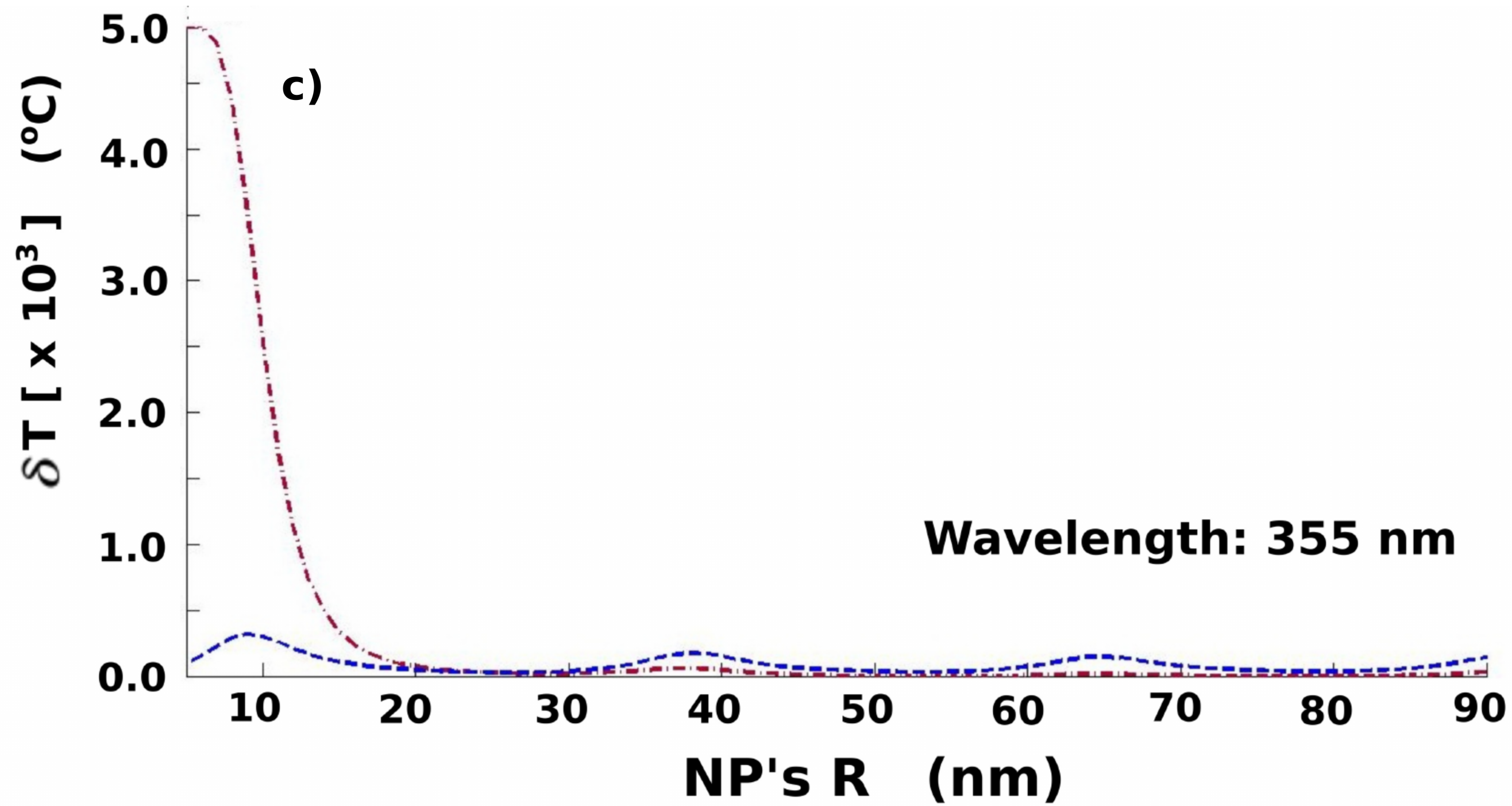}
	\caption {Temperature increase, as function of particle radii and illuminated with (a) 1064 nm, (b) 532 nm and (c) 355 nm. In each case, the line staring at high values correspond to $\delta T$ calculated from (\ref{T}). Line starting at low values is for the estimate using $\Delta T$ from macroscopic description as reported in \cite{SelmkeJOSAA2012, JiangJPCC2013}.}
\end{center} \label{Fig3}	
\end{figure}

\begin{figure}[h]
	\begin{center}
		\includegraphics[scale=0.7]{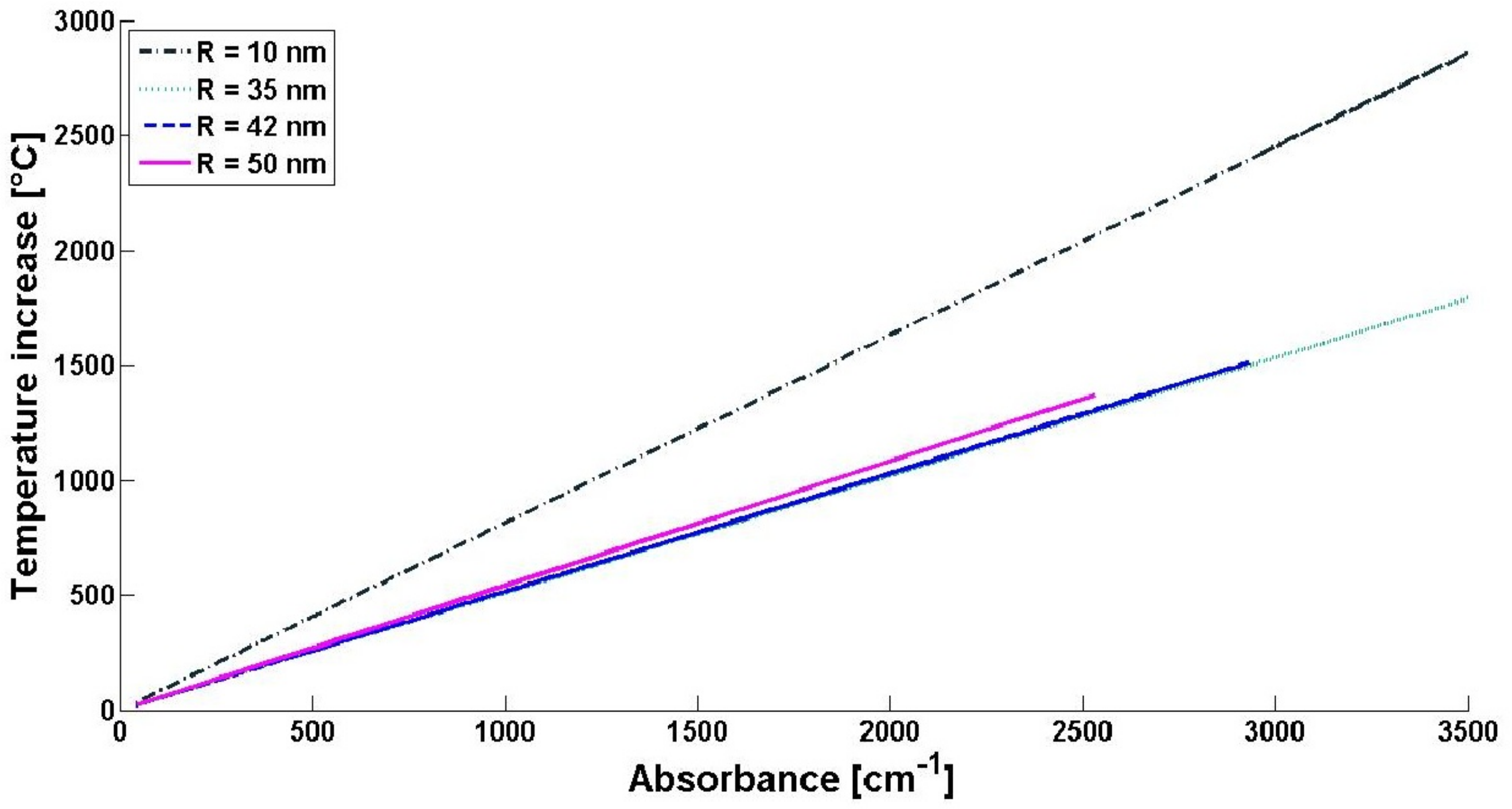} 
		\caption{Temperature change in terms of the optical absorption as function of the excitation wavelength ($\alpha(\lambda)$) for different nanoparticles radii (10, 36, 42 and 50 nm).}\label{Fig4}
	\end{center}
\end{figure}

In Fig. \ref{Fig4} we display the rate of change in temperature in the nanoparticle with respect to optical absorbance in order to show how is the rate of temperature increase as a function of the capacity to absorb radiant  energy, for different nanoparticle sizes.

The model predicts the rate of temperature change to be proportional to the optical absorption,  $\alpha$. Interestingly, as the nanoparticle size reduces, the slope for the change of temperature with optical absorption, is increased. Instead, as the particle size increases, the slope tends to a stable vale.

\section{Discussion}

From both models we obtain a maximum heating when a nanoparticle of 10 nm radius is illuminated with around 347 nm wavelength radiation. Nonetheless for nanoparticles around 20 nm the temperatures reached when they are irradiated by a wavelength among 300 to 400 nm are enough to melt the substrate on which they are supported \cite{Plasmonics, Henley2013}. However, even though the temperature maximum is reached around those wavelengths and sizes, all of the nanoparticles experience an increase in their temperature depending on their diameter, but to a lesser degree. From Fig. (3), we can see that the range of wavelength necessary to heat them up show a displacement to longer wavelengths when the system is larger. Furthermore, as the wavelength decreases, the number of wavelengths at which the nanoparticles will experience a rise in temperature (\textit{hot modes}) increases, even though their peak values tend to decreases with the increase in wavelength. 

We also observe that our model shows a deviation in the peaks when the heating is evaluated as function of size, a behavior that is not present when the radius remain constant. This is explained by the fact that, in both models, the only property that depends on the wavelength is the absorption cross section while the rest of the characteristics evaluated depend on the size of the nanoparticle. 

A final outcome related to the particle size, also extracted form Eq. (\ref{TG_1}), is that since at constant volume $d \rho_e \approx dQ$, thus $dS= \frac{dQ}{T}= \frac{C_v}{T} dT $  can apply. Hence Eq. (\ref{TG_1}), the time derivative of the volume energy density can be equated as

\begin{equation} \label{Q}
\frac{d\rho_e}{dt} =\frac{\kappa\alpha T}{z_1} = C_V \frac{d T}{dt}, 
\end{equation}

where $C_V$ is the volume constant heat capacity. In doing so, one recovers the usual book definition, $d \rho_e / dT = C_V$ and thus  (\ref{Q}) equates as

\begin{equation} \label{C}
\frac{\kappa \alpha }{z_i } dt = C_V \frac{dT}{T} ~~.
\end{equation}

Therefore, since $S= C_V \ln(T_l/T_o)$, after integrating both sides of equation \ref{C}, we get, 

\begin{equation} \label{S}
\frac{\kappa \alpha }{ z_i} t_{\delta} ~ = ~ C_V ln \left(\frac{T_l}{T_o} \right) = S
\end{equation}

After defining $\upsilon_o \equiv z_i/t_{\delta}$, from the last equation we get,

\begin{equation}\label{Cv}
C_V =  \frac{\kappa \alpha }{\upsilon_o ~ln \left(\frac{T_l}{T_o} \right) }   ~ = ~ \frac {S}{ ln\left(\frac{T_l}{T_o} \right)}
\end{equation}

We close this analysis, by noticing: a) from (\ref{Cv}), we get that

\begin{equation} \label{v_o}
\upsilon_o = \frac{\kappa \alpha}{C_V~ln \left(\frac{T_l}{T_o} \right)} ~ =~ \frac{\kappa \alpha}{S} ~~~, \end{equation} 

and b) from arguments in kinetic theory of gases, Peierls  \cite{PeierlsBook2001}  arrived to the near expression:  

\begin{equation} \label{v_s}
\upsilon_s = \frac{\kappa }{c_s \gamma_s \mathcal{L}},
\end{equation}

where $\upsilon_s$ is the sound velocity, $\mathcal{L}$ is a mean free path, $c_s$ is specific heat per unit volume (otherwise the heat capacity of a substance per unit mass), and $\gamma_s$ is a numerical factor. Although from this perspective, (\ref{v_o}) and (\ref{v_s}) keep a remarkable physical and analytic resemblance, to place an equivalence would require a separate analysis. By now, this conclusion suggest the possibility of introducing a `size-scaled' definition of $C $ through its relation to $\kappa$, and in therms of the actual available energy, weighted by the particle size $z_l$ and $\alpha$. As noted by other authors \cite{HeatCapacity}, nano-sized particles exhibit heat capacity value that is somehow different respect of the corresponding bulk value. Thus, one has to be cautious, however, that specific heat is really a unique bulk property and yet at nanoscale, it seems like it require to be reinterpreted. Making it size-dependent would also make it depend of the conversion efficiency of absorbed light into heat; and implicitly, also dependent from the frequency bath in which the system is placed. As noted before, at short scale ($ \leq 20 nm $ particle radii) $\kappa$ tend to be diverge from a constant value \cite{GemmerBook2009}, and thus it may explain the slope for 10 nm in Fig. \ref{Fig4}. However $\kappa \alpha$, seems to balance in a scaled manner as function of the particle size. Clearly extensive experimental verification of these findings are required, and thus properly investigate the $\alpha \kappa /z_l$ interplay.

At this point, we were able to obtain an analytical expression, involving microscopic statistical physics, which describes the changes in temperature experienced by a nanoparticle as function of the illumination wavelength and particle size, prompting the strong exchange of energy through the photothermal effect, for nanoparticles with radius smaller than 42 nm. This, to the best of our knowledge has not reported previously. We propose to describe the effects induced through the discretization of the attainable energy levels. In other words we consider the fact that the nanoscale systems present a modified thermodynamic behavior. In addition, we compare our model with one based on macroscopic properties as previously reported elsewhere in the literature, which as a consistency check shows that, when the size of the gold nanoparticles is smaller than the noted 42 nm threshold, it predicts a greater increase in the temperature with respect to that previously reported in the literature. Mie resonances \cite{JPC2017, ElSayed, HillenbrandNature} are an intrinsic plasmon  property of metallic nanoparticles and therefore the conclusions reported here for the specific case of gold, describe a general form of behavior to be expected (with different characteristics frequencies) for many other metallic nanoparticles.

\section*{Acknowledgments}	
One of the authors MRM acknowledge the scholarship grant from CONACyT. CGS acknowledge the fruitful discussions, criticism and comments from Dr. Edahi Gutierrez-Reyes.


\end{document}